\documentclass{article}

\usepackage{amsthm}
\usepackage{amsmath}
\usepackage{amssymb}
\usepackage{amsfonts}
\usepackage{booktabs} 
\usepackage{graphicx}

\usepackage[margin=2cm]{geometry}

\usepackage{moreverb}

\usepackage[colorlinks,bookmarksopen,bookmarksnumbered,citecolor=red,urlcolor=red]{hyperref}
\usepackage{tabularx}
\usepackage{array}
\usepackage{multirow}

\newcommand\BibTeX{{\rmfamily B\kern-.05em \textsc{i\kern-.025em b}\kern-.08em
T\kern-.1667em\lower.7ex\hbox{E}\kern-.125emX}}

\DeclareMathOperator\Sen{Sen}
\DeclareMathOperator\Spe{Spe}
\DeclareMathOperator\logit{logit}
\DeclareMathOperator\DOR{\omega}
\DeclareMathOperator\lnDOR{\ln\omega}
\DeclareMathOperator\SE{SE}
\DeclareMathOperator\Var{Var}
\DeclareMathOperator\lnt{\ln\vartheta}
\DeclareMathOperator\ESS{ESS}
\DeclareMathOperator\UM{UM}
\DeclareMathOperator\E{E}
\DeclareMathOperator\M{M}
\DeclareMathOperator\B{B}
\DeclareMathOperator\Binom{Binom}
\DeclareMathOperator\T{T}
\DeclareMathOperator\N{N}
\DeclareMathOperator\p{P}

\begin{document}


\title{Testing for Publication Bias in Diagnostic Meta-Analysis: A Simulation Study}

\author{Paul-Christian B\"urkner and Philipp Doebler}

\date{Email: paul.buerkner@gmail.com}

\maketitle



\begin{abstract}
The present study investigates the performance of several statistical tests to detect publication bias in diagnostic meta-analysis by means of simulation. While bivariate models should be used to pool data from primary studies in diagnostic meta-analysis, univariate measures of diagnostic accuracy are preferable for the purpose of detecting publication bias. In contrast to earlier research, which focused solely on the diagnostic odds ratio or its logarithm ($\lnDOR$), the tests are combined with four different univariate measures of diagnostic accuracy.  For each combination of test and univariate measure, both type I error rate and statistical power are examined under diverse conditions. The results indicate that tests based on linear regression or rank correlation cannot be recommended in diagnostic meta-analysis, because type I error rates are either inflated or power is too low, irrespective of the applied univariate measure. In contrast, the combination of trim and fill and $\lnDOR$ has non-inflated or only slightly inflated type I error rates and medium to high power, even under extreme circumstances (at least when the number of studies per meta-analysis is large enough). Therefore, we recommend the application of trim and fill combined with $\lnDOR$ to detect funnel plot asymmetry in diagnostic meta-analysis. Please cite this paper as published in Statistics in Medicine (\url{https://doi.org/10.1002/sim.6177}). \\

Keywords: diagnostic meta-analysis, publication bias, diagnostic odds ratio, trim and fill, simulation study
\end{abstract}


\section{Introduction}
The interest in research synthesis and meta-analysis has rapidly grown over the last few decades. From today's point of view, it is difficult to think of scientific research without the possibility to integrate different findings into one big picture. This is especially true for research on the accuracy of diagnostic tests, because clinical and health policy 
decisions as well as technology development and evaluation in diagnostic medicine have to rely on a good empirical basis \cite{irwig1994, irwig1995,tacconelli2010}. In general, studies of diagnostic accuracy compare the performance of an imperfect diagnostic test with an optimal diagnostic instrument, which is also known as gold standard. Usually, using the gold standard is time-consuming, expensive and / or invasive, so that it can only be applied in certain circumstances. Therefore, alternative tests that are more economic but less accurate have to be evaluated in studies of diagnostic accuracy by comparing them with the gold standard. These studies most often report pairs of \textit{Sensitivity} ($\Sen$) and \emph{Specificity} ($\Spe$), the former being the percentage of correctly diagnosed diseased individuals and the latter
being the percentage of correctly diagnosed healthy individuals. 

As in any other area of research synthesis, results in diagnostic meta-analysis may be biased by 
various effects such as study quality, heterogeneity of the examined populations, 
or, as most commonly cited, \emph{publication bias} (PB) \cite{deeks2005}. PB arises when studies \emph{systematically} remain unpublished
so that they cannot contribute to meta-analysis leading to biased and possible misleading results. "Systematically" comprises that studies are not just missing at random, but that some of the study's characteristics (most often its outcomes) influence the probability of this study being published. 

There are several reasons for researchers to refrain from publishing a study. Some of these reasons occur at random and do not dependent on the outcome of the study (e.g., the retirement of the responsible researcher), so that they do not contribute to the emergence of PB, while other reasons do depend on the outcome: For example, a study's findings may be suppressed by the funding source supporting the research. Besides, the possibly most prominent reason is the non-significance of the results \cite{easterbrook1991}. However, this non-significance rarely applies to diagnostic studies. Usually, findings on test accuracy contain $\Sen$ and $\Spe$ (or univariate measures computed on their basis) together with 95\% confidence intervals, but they do not state a null hypothesis \cite{bossuyt2003,pepe2003}. Hence, in most cases, there is no statistical test that can potentially fail to be significant. Therefore, one might argue that PB is not a problem in diagnostic meta-analysis. Nevertheless, some approaches such as non-inferiority designs (in which a new treatment or diagnostic test is compared with an already established one) are based on significance tests even in diagnostic contexts \cite{cheng2010,li2008,liu2006}. In addition, as studies of diagnostic accuracy are often conducted as part of routine clinical practice, they may be abandoned more easily, which should increase the probability of PB \cite{song2002,tacconelli2010}. Another reason accounting for PB in all types of meta-analysis is that dramatic findings are more likely to occur in primary studies and meta-analyses that are methodically weaker \cite{begg1988}. As scientific journals may prefer such findings, it is plausible that meta-analyses not only tend to overestimate effects because of unpublished studies, but that researchers also base their results on less elaborate literature search. This assumption has been supported by findings of Song \textit{et al.} \cite{song2002} who showed that publication bias was increased in diagnostic meta-analyses in which only few databases were used to find relevant studies. In addition, we share the view of one reviewer of this paper who suspects that potential mechanisms of PB in the diagnostic context will depend on the type of test (e.g. imaging vs. biomarker vs. questionnaire) and the application (e.g. mass screening vs. confirmation of a screen).
In light of the above it becomes clear that the examination and evaluation of PB in diagnostic meta-analysis should not be neglected.

An issue that complicates the detection of PB in diagnostic meta-analysis emerges from different and sometimes implicit cut-off values
used in different studies in order to decide which test scores indicate disease and which do not. The choice of the 
cut-off value depends on the (somewhat subjective) importance of $\Sen$ and $\Spe$ in the respective research context.  
In one study, an individual may be diagnosed as diseased, while in another study with a different cut-off value 
the same person may be diagnosed as healthy. Therefore, even very different pairs of observed $\Sen$ and $\Spe$ might only be caused by  
different cut-off values and may not indicate that the test accuracy itself varied between studies. 
Using univariate effect measures such as the diagnostic odds ratio or its logarithm instead of $\Sen$ and $\Spe$ can reduce the cut-off value problem \cite{glas2003} at the cost of a loss of information. However, the heterogeneity caused by different cut-off values may still be interpreted as PB, even though there is none. In contrast, different cut-off values can also mask an existing PB, so that it cannot be detected. Given all these complications, we evaluated the performance of several statistical tests to detect PB in diagnostic meta-analysis in a comprehensive simulation study.

In general, models of meta-analysis simultaneously have to cope with between and within study
variance. Therefore, random effects models are of critical importance
for many types of meta-analysis, because they are able to separate within study variance from between study variance.
Rutter and Gatsonis \cite{rutter2001} were the first to develop a specific approach of random effects for diagnostic meta-analysis by using hierarchical regression. A few years later, Reitsma \textit{et al.} \cite{reitsma2005} formulated a different, bivariate model to cope with random effects.
It has later been shown by Harbord \textit{et al.} \cite{harbord2007} and likewise and independently by Arends \textit{et al.} \cite{arends2008} that the models of Rutter and Gatsonis \cite{rutter2001} and Reitsma \textit{et al.} \cite{reitsma2005} are very closely related and 
even identical in the absence of covariates. In the present study, the model of Reitsma \textit{et al.} \cite{reitsma2005} was used to sample pairs of $\Sen$ and $\Spe$, as a bivariate model should be used for the analysis of data from primary diagnostic studies \cite{macaskill2010,ma2013,leeflang2008}. However, \emph{pairs} of $\Sen$ and $\Spe$ are difficult to use directly to detect PB, as the common graphical and statistical methods require \textit{univariate measures} (UMs) and to our knowledge bivariate tests have not been developed so far \cite{ma2013}. In our study, four UMs that were computed on the basis of $\Sen$ and $\Spe$ were each combined with every applied statistical test. The performance of these combinations was evaluated by means of simulation.

Section 2 reviews the bivariate model, different UMs used in diagnostic studies, as well as existing statistical tests to detect PB. 
The simulation process and its results are described in detail in Section 3 and 4.  
In Section 5, the findings, implications and limitations of the present study are summarized and discussed.

\section{Theory and methods}
\subsection{Bivariate Sampling Model}
As mentioned above, the bivariate model of Reitsma \textit{et al.} \cite{reitsma2005} was used as a sampling model for our simulations. Let $k$ be the number of studies summarized in the meta-analysis and $i = 1,...,k$.  
The basic assumption is that the true pair of $\logit(\Sen_{i})=\theta_{A,i}$ and $\logit(1-\Spe_{i})=\theta_{B,i}$
is bivariate normally distributed with mean $\mu=(\theta_{A}, \; \theta_{B})^{\text{T}}$ and between study covariance matrix $\Sigma$:
\begin{equation}
\begin{pmatrix} \theta_{A,i} \\ \theta_{B,i} \end{pmatrix}  \sim  \N(\mu, \; \Sigma) \quad \text{with} \quad  \Sigma= \begin{pmatrix} \sigma^2_{A} & \sigma_{AB} \\
        \sigma_{AB} & \sigma^2_{B} \end{pmatrix}.
\end{equation} 
Despite the fact that this model explicitly considers the bivariate nature of diagnostic data, it has the advantage that it directly accounts for the covariance between $\Sen$ and $\Spe$ through $\sigma_{AB}$. 

\subsection{Univariate effect measures}
Comparing diagnostic tests only on the basis of $\Sen$ and $\Spe$ can be problematic, as one test may have a higher $\Sen$, while the other test may have a higher $\Spe$. If this is the case, it will be difficult to decide, which test is to be preferred unless one indicator is obviously more important. The advantage of UMs combining the information on $\Sen$ and $\Spe$ is that different diagnostic tests are always comparable, at the expense of losing some information due to the reduction to one single measure. With respect to the present study, UMs are \textit{only} of importance, as statistical tests to detect PB require UMs instead of pairs of $\Sen$ and $\Spe$ \cite{ma2013}. In the following, four different UMs are introduced.     
The notation is given in Table~\ref{2x2}.

\begin{table}[hbtp]
\centering
\caption{Data from a diagnostic study in a 2$\times$2 table.}
\label{2x2}
\renewcommand{\arraystretch}{1.5}
\begin{tabular}{ccccc} \toprule
 &  & \multicolumn{3}{c}{Diagnostic Test}  \\ \cmidrule{3-5}
 &  & Positive & Negative & Total \\ \midrule
 \multirow{2}{*}{Gold standard} & Positive & $x$ & $w$ & $n_{1}$ \\
 & Negative & $y$ & $z$ & $n_{2}$ \\ 
 & Total   & $m_{1}$ & $m_{2}$ & $N$  \\ \bottomrule	
\end{tabular} 
\end{table}
 
The probably most prominent UM in the context of diagnostic studies is the diagnostic odds ratio ($\DOR$; or its natural logarithm $\lnDOR$). In general, one computes the \textit{standard error} ($\SE$) and the related confidence interval of $\lnDOR$ as it is approximately normally distributed unless the observed frequencies (i.e. $w$, $x$, $y$ and $z$) are too small. It holds that:
\begin{equation}
  \lnDOR := \ln\left(\frac{xz}{yw}\right) \quad
\end{equation}
with 
\begin{equation}
  \SE(\lnDOR) = \sqrt{\frac{1}{x}+\frac{1}{y}+\frac{1}{w}+\frac{1}{z}}.
\end{equation} 
The range of $\lnDOR$ is $-\infty$ to $+\infty$, with higher values standing for higher test accuracy and values lower than zero representing tests that are worse than guessing. In case of zero cells in the underlying 2$\times$2 table, it is generally recommended to add $0.5$ to each observed frequency in order to calculate an approximation of $\lnDOR$ \cite{glas2003,littenberg1993}, although there are other methods that may be less biased \cite{sweeting2004}. Also, $\lnDOR$ is not affected by unequal sample sizes \cite{haddock1998}. Importantly, tests to detect PB in diagnostic meta-analysis are almost without exception based on $\DOR$ or $\lnDOR$ while other UMs have not been discussed so far (see Schwarzer \textit{et al.} \cite{schwarzer2002} for the only exception known to the authors, in which the risk ratio was examined and performed worse then $\lnDOR$. Therefore, the risk ratio was not included in our simulations). Among other aims, the present study investigated whether this focus on the diagnostic odds ratio can be regarded as justified.  

Another UM proposed by Le \cite{le2006} is derived from a model, which describes the relationship between $\Sen$ and $\Spe$ using the Lehmann family:
\begin{equation} \Sen = (1-\Spe)^\vartheta \quad  \text{with} \quad 0 < \vartheta \leq 1. \end{equation}
This model allows to approximate the receiver operating characteristic (ROC), which is often used to summarize 
diagnostic results that are continuous (i.e. not binary \cite{holling2012a,holling2012b,le2006}). Solving for the natural logarithm of $\vartheta$ reveals:
\begin{equation} 
  \lnt := \ln\left(\frac{\ln(x)-\ln(n_{1})}{\ln(y)-\ln(n_{2})}\right) 
	\end{equation}
with the standard error
\begin{equation}
  \SE(\lnt) = \sqrt{ \dfrac{\frac{1}{x}-\frac{1}{n_{1}}}{(\ln(x)-\ln(n_{1}))^{2}} + \dfrac{\frac{1}{y}-  \frac{1}{n_{2}}}{(\ln(y)-\ln(n_{2}))^{2}}}.
\end{equation}
Similar to $\lnDOR$, $\lnt$ is not affected by unequal sample sizes and ranges from $-\infty$ to $+\infty$ with values greater than zero representing tests that are worse than guessing. Thus, in contrast to $\lnDOR$, higher values represent lower test accuracy. However, some tests for PB require higher values to be associated with higher test accuracy. Therefore, -$\lnt$ was used in our simulations instead.

Youden's Index \cite{youden1950}, in our notation written as 
\begin{equation} 
Y := \frac{x}{n_{1}} + \frac{z}{n_{2}} - 1, 
\end{equation}
is yet another UM used in diagnostic studies. The value of $Y$ is fairly constant, even if $\Sen$ and $\Spe$ are varying due to different cut-off scores \cite{bohning2008}. Also, $Y$ appears to be a better measure to choose an appropriate cut-off value than $\DOR$ \cite{bohning2011}. These are important properties when dealing with the cut-off value problem. Moreover, $Y$ is very simple to calculate, it is unaffected by unequal sample sizes, and its values range from -1 to 1. Assuming that a diagnostic test is equal or better than a random decision, 
$Y$ only ranges from 0 to 1 (the higher $Y$ the better the test). It holds that: 
\begin{equation}
\SE(Y) = \sqrt{\dfrac{\frac{x}{n_{1}}(1-\frac{x}{n_{1}})}{n_{1}} + \dfrac{\frac{y}{n_{2}}(1-\frac{y}{n_{2}})}{n_{2}} }.
\end{equation}
Due to the simplicity of $Y$ and its usefulness when choosing an appropriate cut-off value, it has frequently been applied in diagnostic studies. With respect to meta-analysis, B\"ohning \textit{et al.} \cite{bohning2008} proposed an overall estimator for the performance of a diagnostic test based on $Y$, including the possibility to add mixtures to cope with unobserved heterogeneity. 

A measure that behaves somewhat similar to $Y$ is Cohen's Kappa \cite{cohen1960}:
\begin{equation}
K := \frac{2(xz-yw)}{n_{1}m_{2}+n_{2}m_{1}}.
\end{equation} 
Although it can be applied in diagnostic contexts \cite{kraemer2002,bloch1997}, $K$ is generally known as a measure of the agreement between two or (if generalized) more raters \cite{fleiss1971}, but not as a measure of diagnostic accuracy. Accordingly, when compared to the other measures mentioned above, $K$ is less often applied in diagnostic studies. Calculating $\SE(K)$ is quite complex. The SE used in the present study was proposed by Fleiss \textit{et al.} \cite{fleiss1969} (under a multinomial assumption) and is the most common: 
\begin{equation}
SE(K) = \dfrac{\sqrt{A+B-C}}{(1-(n_{1}m_{1}+n_{2}m_{2})/N^2)\sqrt{N}}
\end{equation} 
where 
  \begin{eqnarray}
  & & \nonumber A := \frac{1}{N^3}(x(N-(n_{1}+m_{1}(1-K))^2+w(N-(n_{2}+m_{2}(1-K))^2), \\ 
  & & B := \frac{1}{N^3}(1-K)^2(w(n_{2}+m_{1})^2+y(n_{1}+m_{2})^2), \\
  & &  \nonumber C := (K - (n_{1}m_{1}+n_{2}m_{2})(1-K)/N^2)^2.
	\end{eqnarray}
When sample sizes are equal in both groups, $K$ is identical to $Y$ with $\SE(K)$ and $\SE(Y)$ being very similar. However, when sample sizes are unequal, both measures differ, because in contrast to the other UMs, 
$K$ depends on the sample sizes per group. For sample distributions of the UMs in the absence and presence of PB see Figure~\ref{densities} in Section 3. 

\subsection{Detecting publication bias using funnel plots}

\begin{figure}[btp] 
  \centering
    \includegraphics[scale=0.70]{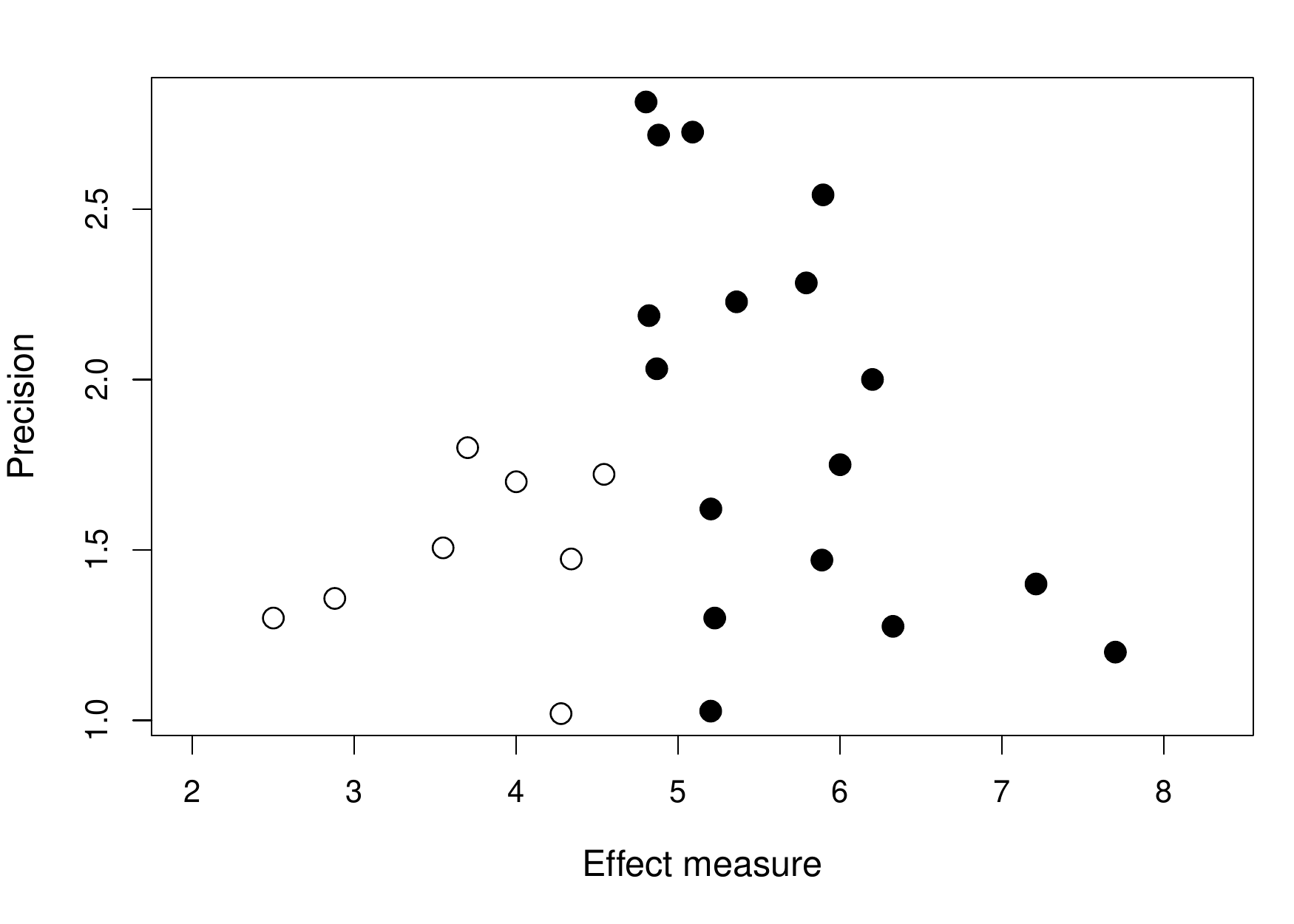}
	\caption[3pt]{Hypothetical funnel plot illustrating published studies (solid points) and unpublished studies (white points). As the unpublished studies cannot be considered for meta-analysis, the funnel emerged from the solid points is asymmetric and hints at the presence of PB.}
\label{funnel}
\end{figure}

A basic graphical method to detect PB are funnel plots, in which effect 
measures (such as those mentioned above) are plotted against the total sample size $N$ or the precision $\SE^{-1}$ of the effect measure \cite{richard1984}. The shape of the funnel plot visualizes whether PB is existent or not. If not existent, the points will form a symmetrical funnel around an overall estimated effect. More precisely, studies with low $N$ or high SE should spread more around the overall effect, while studies with high $N$ or low SE should be closer to it \cite{deeks2005}. However, if small studies with low or non-significant effects remain unpublished, the funnel plot should appear asymmetrical, which may lead to the conclusion that PB is existent. A sample funnel plot of a hypothetical meta-analysis in which PB is present is depicted in Figure~\ref{funnel}.

Notably, funnel plots and statistical tests cannot discriminate between PB and other sources that cause funnel plot asymmetry \cite{rothstein2006}. The cut-off value problem in diagnostic meta-analysis further complicates the detection of PB. Since funnel plots of $\Sen$ and $\Spe$ do particularly suffer from this cut-off value problem, UMs should be used instead. Glas \textit{et al.} \cite{glas2003} showed, that the DOR might depend on the cut-off (as implied by the model of \cite{rutter2001}), but is reasonably constant nevertheless. In this sense, UMs alleviate the cut-off value problem. Moreover, three dimensional funnel plots displaying $\Sen$ and $\Spe$ at the same time are possible, but they may be too difficult for reasonable interpretation. Yet even without these complications, funnel plots cannot be seen as a reliable method to detect PB \cite{terrin2005}. Therefore, statistical tests are needed.  

\subsection{Detecting publication bias using statistical tests}

The present study examined the performance of several statistical tests that were developed to detect PB by means of simulation. All of these tests were combined with each of the four UMs (i.e. $\lnDOR$, $\lnt$, 
$Y$, and $K$). The tests do all have in common that they are more or less based on funnel plots. Contrary to funnel plots, however, formal instead of visual criteria decide if PB is present. When PB is genuinely existent in diagnostic meta-analysis, studies that found low diagnostic accuracy remain unpublished and cannot be considered for meta-analysis. Therefore, by using UMs with higher values corresponding to higher diagnostic accuracy as in the present study, missing studies should appear on the left part of the funnel. Conclusively, it seems reasonable to apply tests for detecting PB with one-sided (instead of two-sided) hypotheses. 

Concerning the performance of a statistical test, two types of errors are of importance. A type I error ($e_{1}$) occurs when the test is significant, even though there is no PB. In contrast, a type II error ($e_{2}$) occurs, when there is PB but the test fails to be significant. Let $\alpha$ be the nominal $\alpha$-level of a test and $\p(x)$ the probability of an event $x$. A test is called conservative when $\p(e_{1}) < \alpha$, whereas it is called liberal (or anti-conservative) when $\p(e_{1}) > \alpha$. The power of tests is generally defined as $1-\p(e_{2})$. Taken together, statistical tests are considered as good when they are not liberal and have a high power.

\subsubsection{Egger regression}
Egger \textit{et al.} \cite{egger1997} used a linear regression approach to detect PB. They suggested to regress a UM divided by its SE as dependent variable on the precision of that UM as independent variable: 
\begin{equation}
\UM \cdot \SE^{-1} = b_{0} + b_{1}\SE^{-1} + \, \varepsilon,
\end{equation}
where $\varepsilon$ represents the error arising from the prediction. In case of no PB, small studies are expected to be close to zero on both axes (due to their high SE), whereas larger studies have a high precision and are thus expected to deviate from zero. In this case, the regression line is assumed to go through the origin, so that $b_{0}$ does not differ significantly from zero. However, if PB is existent, most small studies will have relatively high UMs and $b_{0}$ is thus assumed to be significantly greater than zero.
The Egger regression can be altered by replacing $\SE^{-1}$ in (12) by the total sample size $N$.
In addition, each study can be weighted by the inverse of the variance under a fixed or random effects assumption to control for possible heteroscedasticity (c.f. \cite{thompson1999}). These variations of the Egger regression have in common that they are mostly too liberal when combined with (diagnostic) odds ratio or its logarithm \cite{deeks2005,macaskill2001,schwarzer2002}. For this reason, Harbord \textit{et al.} \cite{harbord2006} proposed another variation of Egger's regression specifically designed for meta-analysis of binary outcomes, which uses the efficient score of an UM and the variance of the efficient score (for details see \cite{harbord2006}). Furthermore, there are alternative regression tests for meta-analysis of binary outcomes, which utilize the arcsine transformation in order to hold the nominal $\alpha$-level \cite{rucker2008}.

\subsubsection{Macaskill regression}
Macaskill \textit{et al.} \cite{macaskill2001} introduced a different regression approach to detect PB. They suggested to regress a UM as dependent variable on the total sample size $N$ as independent variable and weight the studies by the inverse of the variance:
\begin{equation}
\UM = b_{0} + b_{1}N + \, \varepsilon.
\end{equation}
In case of no PB, the values of the UM should be constant across different sample sizes and $b_{1}$ is thus assumed to be close to zero. However, when PB is present, small studies should have greater UMs on average, which causes $b_{1}$ to be below zero. Deeks \textit{et al.} \cite{deeks2005} suggested to replace $N$ by $1/\sqrt{\ESS}$ (i.e. the \textit{effective sample size} calculated as $\ESS = (4n_{1}n_{2})/(n_{1}+n_{2})$) and weight the studies by $\ESS$ to achieve a better performing test, especially when there are only few diseased persons in the sample. A similar approach was suggested by Peters \textit{et al.} \cite{peters2006}, who replaced $N$ by $1/N$ (for details on the weighting see \cite{peters2006,peters2005}). Note that, when $\ESS$ or $1/N$ are used, $b_{1} > 0$ has to be tested. Although it seems reasonable to further variate the Macaskill regression by replacing $N$ by $\SE$, Sterne \textit{et al.} \cite{sterne2000} demonstrated that the resulting test is equivalent to Egger's regression.  
Notably, in the approach of Egger \textit{et al.} \cite{egger1997} $b_{0}$ is tested for significance in order to detect PB, whereas the method of Macaskill \textit{et al.} \cite{macaskill2001} uses $b_{1}$.  

\subsubsection{Begg's rank correlation}
Begg and Mazumdar \cite{begg1994} proposed a non-parametric rank correlation method  to detect PB, which is based on Kendall's tau ($\tau$; \cite{kendall1938}). Let $t_{i}$ be the effect measure and $\Var_{i} := \Var(t_{i}) = \SE(t_{i})^2$  the related variance of study $i$ in the meta-analysis. Then one calculates:
\begin{equation}
t_{i}^{*} := (t_{i} - \bar{t}) / \SE_{i}
\end{equation} 
with  
\begin{equation}
\bar{t} := (\sum \Var_{i}^{-1}t_{i}) / \sum \Var_{i}^{-1}
\end{equation} 
being the common fixed effects estimator of the overall effect. It is then tested whether $t_{i}^{*}$ and $\Var_{i}$ are significantly associated (i.e. if $\tau$ differs significantly from zero). In the absence of PB, the variance should be independent of the effect measure and thus $\tau$ is assumed to be close to zero. In the presence of PB, some small studies will have inflated effect measures and hence large values $t_{i}^{*}$.  For that reason, a $\tau$ that is significantly greater than zero indicates the presence of PB. It has been shown that Begg's rank correlation is a little conservative and has less power than Egger's regression when used with (diagnostic) odds ratio or its logarithm \cite{deeks2005,macaskill2001,sterne2000}. Possible variations of Begg's rank correlation can be obtained by replacing $\Var$ by $N^{-1}$ or $\ESS^{-1}$, respectively \cite{deeks2005}. Further variation specifically designed for binary data were proposed by Schwarzer \textit{et al.} \cite{schwarzer2007} and R\"ucker et al. \textit{et al.} \cite{rucker2008}. 
 
\subsubsection{Trim and fill}
Trim and fill is another non-parametric method to detect PB and was developed by Duval und Tweedie \cite{duval2000}. It is based on the idea that there are $k$ studies present in the meta-analysis and $k_{0}$ studies missing due to PB, which implies an asymmetrical funnel. With respect to trim and fill, funnel plots are applied with an UM on the $x$-axis and its precision (or alternatively $N$) on the $y$-axis. If we assume to know the true overall effect $\Theta$, we will be able to estimate $k_{0}$. Let $t_{i}^{+} := t_{i} - \Theta$, $r_{i}^{+}$ be the rank of $|t_{i}^{+}|$ ($ 1 \leq r_{i}^{+} \leq k$) and $\gamma^{+} \geq 0$ be the rightmost run of ranks associated with positive values of $t_{i}^{+}$. Then $k_{0}$ can be estimated by
\begin{equation}
R := \gamma^{+} - 1 
\end{equation}
or 
\begin{equation}
L := \frac{4\sum_{t_{i}^{+} > 0}r_{i}^{+} - k(k+1)}{2k-1}.
\end{equation}
In addition, the authors of trim and fill proposed a third estimator of the number of missing studies, but did not recommend its usage \cite{duval2000}. Therefore, this estimator was not considered in the present study.
 
As $\Theta$ is usually unknown, it has to be estimated as well. Using an easily computed iterative algorithm (discussed in detail in \cite{duval2000}), one arrives at a random effects estimator  
$\hat{\Theta}$ that is used in the formula above. In the absence of PB (i.e. $k_{0} = 0$), the approximate distributions of $R$ and $L$ are known. Thus, $R$ and $L$ can be tested for significance to decide whether PB is present or not. As $\hat{\Theta}$ can be seen as corrected for PB \cite{duval2000}, trim and fill is not only a method to test for PB but it also offers to correct the overall estimator (for further discussion see \cite{duval2000,peters2007}) . Note that trim and fill can only be applied with one-sided hypotheses as apposed to all other tests mentioned above.

In literature, several alternative methods other than trim and fill were proposed to estimate the number of missing studies \cite{dear1992,givens1997,hedges1992}, but they are far more complex than any of the methods discussed in this section. As a method's success not only rests upon its performance, but also upon its simplicity and applicability in practice, these complex alternative methods were not considered in the present study.

All other statistical tests discussed in this section were included in our simulation and each of them was combined with every of the four UMs (and with every of the accuracy measures $\SE$, $N$, and $\ESS$) as far as possible and feasible. In view of the large number of tests for PB simulated in our study, it makes sense to introduce a short form at least for those combinations, which are frequently mentioned in our results and discussion (see Table~\ref{shortforms}; for notational convenience, UMs are denoted as $t$, whereas measures of its accuracy are denoted as $v$). Note that not all test variations have got a short form in order to keep the number of notations at an acceptable level.    

\begin{table}[hbtp]
\centering
\caption{Short forms of the tests to detect PB.}
\label{shortforms}
\renewcommand{\arraystretch}{1.5}
\begin{tabular}{lll} \hline
    Test          & Short form           & Additional notations \tabularnewline \hline
    Egger         & $\E(t,v,w)$    & $w$: weights of each study \tabularnewline
    Macaskill     & $\M(t,v)$    &  \tabularnewline
    Begg          &   $\B(t,v)$         &      \tabularnewline
    Trim and fill &  $\T(t,v,m)$ & $m$: estimator of $k_{0}$ \tabularnewline \hline
\end{tabular} \tabularnewline
\end{table}
For example, if $Y$ is used in the original version of Egger's regression with every study being equally weighted (so that the weight argument can be suppressed), the resulting test will be shortened to 
\begin{equation}
E(Y,\SE).
\end{equation}
If $\lnt$ is plotted against $N$ and tested with trim and fill while $k_{0}$ is estimated by the statistic $R$ stated in (16), we will write
\begin{equation}
T(\lnt,N,R).
\end{equation} 

\section{Simulations}

In order to evaluate the performance of the tests to detect PB discussed above, diagnostic data were simulated with and without PB. In addition, several other parameters such as the number of studies per meta-analysis $k$ or the mean quality $\mu$ of the diagnostic test were systematically varied. In the present simulation, $k$ took on values of $k = 30$ (many studies) or $k=10$ (few studies). The total sample size $N$ of each of the $k$ studies was randomly sampled from a discrete uniform distribution and varied between $N = 50$ and $N = 1000$. The prevalence $\pi$ (i.e. the rate of diseased individuals) took on values of $\pi = 0.5$ ($n_{1} = n_{2} = 0.5N$; balanced) or $\pi = 0.2$ ($n_{1} = 0.2N$, $n_{2} = 0.8N$; unbalanced). The resulting $n_{1}$ and $n_{2}$ were always rounded to nearest integer. Simulations were implemented in \texttt{R} \cite{Rcore,Rcpp,meta} with parts of the program coded in \verb!C++! in order to decrease the simulations' duration.

In contrast to earlier simulation models concerning diagnostic accuracy (c.f. \cite{deeks2005}), for each study the true logits of $\Sen$ and $1-\Spe$ were directly sampled using the model of Reitsma \textit{et al.} \cite{reitsma2005} stated in (1), to consider the bivariate structure of diagnostic data. Both, $\mu$ and $\Sigma$ were systematically varied (see Table~\ref{summaryP} for the exact values). With regard to the true mean $\mu$ of the logit diagnostic accuracies, four different values were selected representing a random decision, a diagnostic test with overall low accuracy, a test with overall high accuracy, or a test with only high $\Sen$. The alternative of a test with only high $\Spe$ was not considered, because its results were assumed to be similar to the high $\Sen$ condition due to symmetry. 
Regarding the between study variance $\Sigma$, which is, among others, assumed to be caused by different cut-off values, three different values were selected representing a fixed effects assumption (in which no between study heterogeneity is present) or small / large random effects respectively. The values of $\Sigma$ in the random effects conditions were chosen to be similar to real data of diagnostic meta-analysis.

Each point sampled from (1) can unambiguously be transformed back to its related $\Sen$ and $1-\Spe$ by taking the inverse of the logit. After the true pair of $\Sen_{i}$ and $1-\Spe_{i}$ of study $i$ was sampled, a binomial error was added in order to include the error of measurement:
\begin{equation}
x_{i} \sim \Binom(\Sen_{i},n_{1,i}) \quad \text{and} \quad y_{i} \sim \Binom(1-\Spe_{i},n_{2,i}).
\end{equation}

Taken together, these values allowed to fill in the diagnostic 2$\times$2 table (see Table~\ref{2x2}) for each study and thus, every UM and its respective SE could be calculated on that basis. 

\subsection{Introducing publication bias}
 
In general, PB is modeled in such a way that the probability of a study to be published depends on its effect measure or its $p$-value \cite{begg1994,deeks2005,macaskill2001}. The lower the effect measure 
(or the higher the $p$-value), the lower the probability of the study to be published. For the present simulations, two different methods to introduce PB were considered. 

The first method was similar to those proposed in literature. It was based on the idea to exclude $l$ studies with the lowest Youden index from the meta-analysis. In order to finally arrive at $k$ studies (more precisely at $k = 30$ or $k = 10$), $k + l$ studies were simulated in a first step. In a second step, those $l$ studies with the lowest $Y$ were excluded. In the following, this procedure is referred to as \textit{selection}. As not only the existence of PB but also its strength was varied, $l$ took on values rounded to the nearest integer of $l = 0.2k$ (small PB) or $l = 0.4k$ (large PB). Contrary to methods proposed in literature, the decision to exclude a study from the meta-analysis depended on the outcomes of all other studies. 

In the second method, no studies were excluded from meta-analysis to introduce PB. Instead, it was assumed that some studies do report systematically higher diagnostic accuracy than other studies. For instance, as developers of diagnostic tests are usually interested in presenting their test in a good light, they may choose certain experimental settings in order to obtain (possibly unrealistic) high diagnostic accuracies. To model this assumption, two thirds of the $k$ studies were sampled from $\N(\mu,\Sigma)$, whereas one third was sampled from $\N(\mu + \eta,\Sigma)$. Here, $\eta$ describes the strength of the PB. In the following, this method is referred to as \textit{mixture}. In the present simulations, $\eta$ took on values of $\eta = (0.75, \; \text{-}0.75)^\text{T}$ (small PB) or $\eta = (1.25, \; \text{-}1.25)^\text{T}$ (large PB). Though it may be argued that this kind of systematic heterogeneity can be addressed by adding moderators to the meta-analysis, well performing tests to detect PB should be able to detect this type of bias. A summary of all varied parameters is provided by Table~\ref{summaryP}.

\begin{table}[hbtp]
\centering
\caption{Summary of all parameters varied.}
\renewcommand{\arraystretch}{1.5}
\begin{tabular}{ccccm{5cm}} \hline
    $\mu$     & $\Sigma$     &  $k$    & $\pi$ &  bias \tabularnewline \hline \addlinespace
		
		random decision:     &  fixed effects:    &  few studies:    & balanced: &  none: \tabularnewline \addlinespace
    $ \mu = \begin{pmatrix} \hphantom{.}0\hphantom{.} \\ \hphantom{.}0\hphantom{.} \end{pmatrix}$  &  $\Sigma = \begin{pmatrix} \hphantom{.}0\hphantom{.} & \hphantom{.}0\hphantom{.} \\
        \hphantom{.}0\hphantom{.} & \hphantom{.}0\hphantom{.} \end{pmatrix}$ & $k = 10 $ & $\pi = 0.5 $   &  no PB \tabularnewline \addlinespace
				
		low accuracy:     &  small random effects:    &  many studies:    & unbalanced: &  selection small: \tabularnewline \addlinespace
    $\mu = \begin{pmatrix} 1 \\ -1 \end{pmatrix}$ & $\Sigma = \begin{pmatrix} 0.5 & 0.3 \\
        0.3 & 0.5 \end{pmatrix}$ & $k = 30 $ & $\pi = 0.2 $ &  removal of $0.2k$ studies with the lowest Youden Index  \tabularnewline \addlinespace
				
		high accuracy:     &  large random effects:    &      &  &  selection large: \tabularnewline \addlinespace
    $\mu = \begin{pmatrix} 2 \\ -2 \end{pmatrix}$ & $\Sigma = \begin{pmatrix} 1 & 0.5 \\
        0.5 & 1 \end{pmatrix}$ &         &            & removal of $0.4k$ studies with the lowest Youden Index \tabularnewline \addlinespace
				
		high sensitivity:     &    &      &  &  mixture small: \tabularnewline \addlinespace
    $\mu = \begin{pmatrix} 2 \\ -1 \end{pmatrix}$ &              &         &            & mixture of two distributions with the mean of the second shifted by $(0.75, \; \text{-}0.75)^\text{T}$  
		\tabularnewline \addlinespace
		
		          &              &         &            & mixture large: \tabularnewline \addlinespace
              &              &         &            & mixture of two distributions with the mean of the second shifted by $(1.25, \; \text{-}1.25)^\text{T}$ \tabularnewline \hline
\end{tabular} \tabularnewline
\label{summaryP}
\end{table}
 
\begin{figure}[btp] 
  \centering
    \includegraphics[scale=0.45]{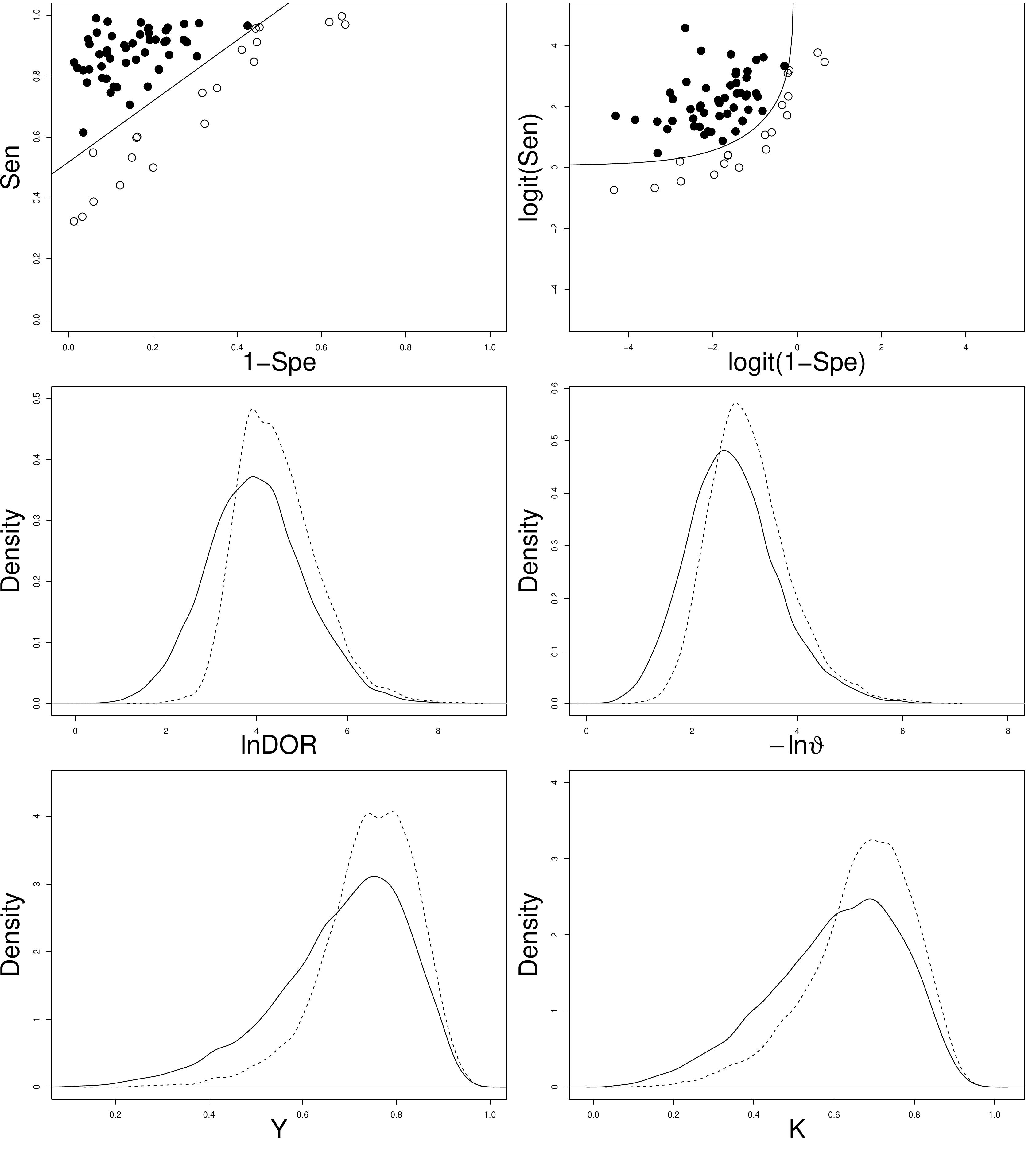}
	\caption[3pt]{In the first row, 70 simulated studies (under the condition of high accuracy, large random effects, and unbalanced prevalence) are plotted at the ROC and logit ROC Space. Solid points symbolize published studies and white points symbolize unpublished studies. The lines depict the cut-off criterion that decides which studies are treated as unpublished. In the second and third row, the related smoothed densities are illustrated in the absence of PB (solid line) and its presence (dotted line).}
	\label{densities}
\end{figure}

To get an impression on how the UMs are distributed when PB is either existent or not, Figure~\ref{densities} illustrates the selection method and the resulting smoothed densities. One realizes that the UMs may be asymmetrically distributed even in the absence of PB (the amount of asymmetry dependents on the parameter values in the underlying simulation).

In the present simulations, all combinations of the values for the parameters were taken into account, resulting in 240 unique combinations. For each of these combinations, 10000 meta-analyses were simulated, so that the accuracy of the results was ensured up to the third decimal place. Each of these meta-analyses was tested for PB with $\alpha = 10\%$ (which was recommended for tests to detect PB; \cite{begg1994,egger1997,macaskill2001}). It was recorded which tests did indicate PB and which tests did not. 
  
\section{Results}
Although all combinations of UMs and tests to detect PB were simulated, only those findings of central importance are discussed in this section. In a first step it is discussed, which UM performs best when combined with the common versions of Egger, Begg, and trim and fill. In a second step, the performance of a broader variety of tests was each combined with the best UM resulting from the first step. 
\subsection{Comparison of the UMs}
\begin{figure}[btp] 
  \centering
    \includegraphics[scale=0.75]{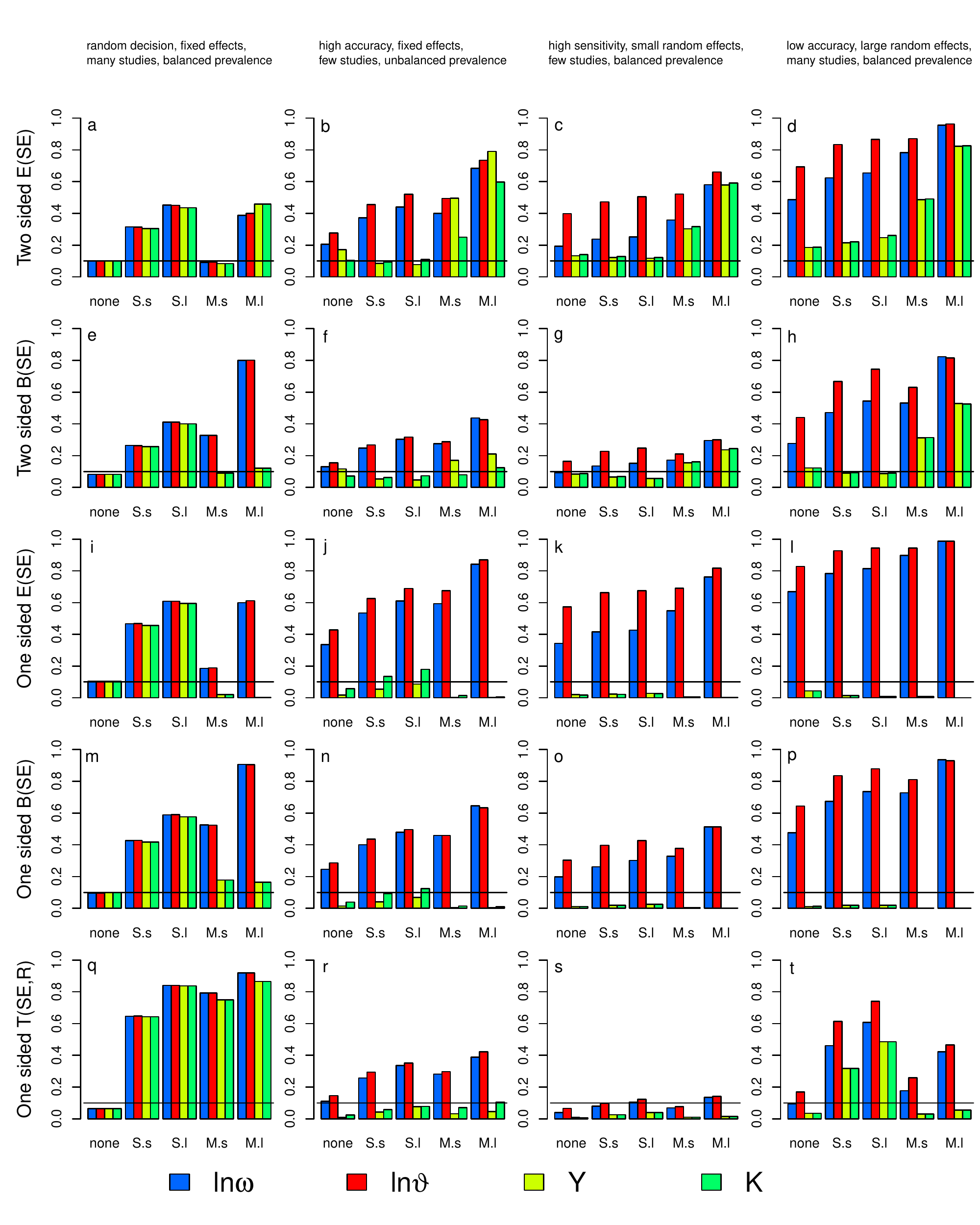}
	\caption[3pt]{Type I error rates and statistical power for the comparison of the UMs. The conditions pictured in this Figure were selected for being most representative. Abbreviations: S.s = selection with small PB, S.l = selection with large PB, M.s = mixture with small PB, M.l = mixture with large PB.}
	\label{step1}
\end{figure}

In case of random decisions and fixed effects, all common tests combined with each UM nearly held the $\alpha$-level of $10\%$, irrespective of the other parameters (see Figure~\ref{step1}). However, the more $\mu$ and $\Sigma$ differed from zero, the more liberal most tests combined with $\lnDOR$ or $\lnt$ were (for instance see plots j, k, l, and p in Figure~\ref{step1}). In contrast, one-sided tests combined with $Y$ and $K$ mostly had zero type I error rates and zero power when the diagnostic test was better than a random decision and when random effects were present (see third to fifth row in Figure~\ref{step1}). As this was not found for two-sided tests, further explanations follow in Section 5. For better comparison of the UMs, Figure~\ref{step1} does also illustrate the results of two-sided tests. 

The UMs $Y$ and $K$ were not considered to be adequate for detecting PB for two reasons. First, they cannot reasonably be used with one-sided tests and second, they rarely performed better than $\lnDOR$ or $\lnt$ when two-sided tests were applied. Although $\lnt$ had higher power than $\lnDOR$ when used with $\E(\SE)$, $\B(\SE)$, or $\T(\SE,R)$, it was also more liberal. More precisely, the greater the power difference between $\lnt$ and $\lnDOR$, the greater the type I error rate difference between $\lnt$ and $\lnDOR$. Importantly, $\T(\lnDOR,\SE,R)$ had non-inflated or only slightly inflated $\alpha$-levels (see fifth row of Figure~\ref{step1}). For these reasons, $\lnDOR$ indeed seemed to the best UM for detecting PB in diagnostic meta-analysis. Therefore its frequent application for research purposes can be considered as justified.

\subsection{Comparison of the tests}
The findings suggest to concentrate on $\lnDOR$, so that this section only focuses on tests combined with $\lnDOR$. In the following, the UM argument in the short forms is suppressed for notational convenience. As can be seen in Figure~\ref{step1}, E(SE) as well as B(SE) had highly inflated $\alpha$-levels when the diagnostic test was better than a random decision and when random effects were present, especially in case of many studies included in the meta-analysis. The highly inflated $\alpha$-levels were also found when Egger's regression was weighted by the inverse of the variance. Furthermore, the variation of Egger's regression proposed by Harbord \textit{et al.} \cite{harbord2006} and the alternative rank correlation test proposed by Schwarzer \textit{et al.} \cite{schwarzer2007} (both not shown in the Figures) were less liberal and much less powerful than E(SE) or B(SE), but still had inflated $\alpha$-levels in case of random effects.

\begin{figure}[btp] 
  \centering
    \includegraphics[scale=0.75]{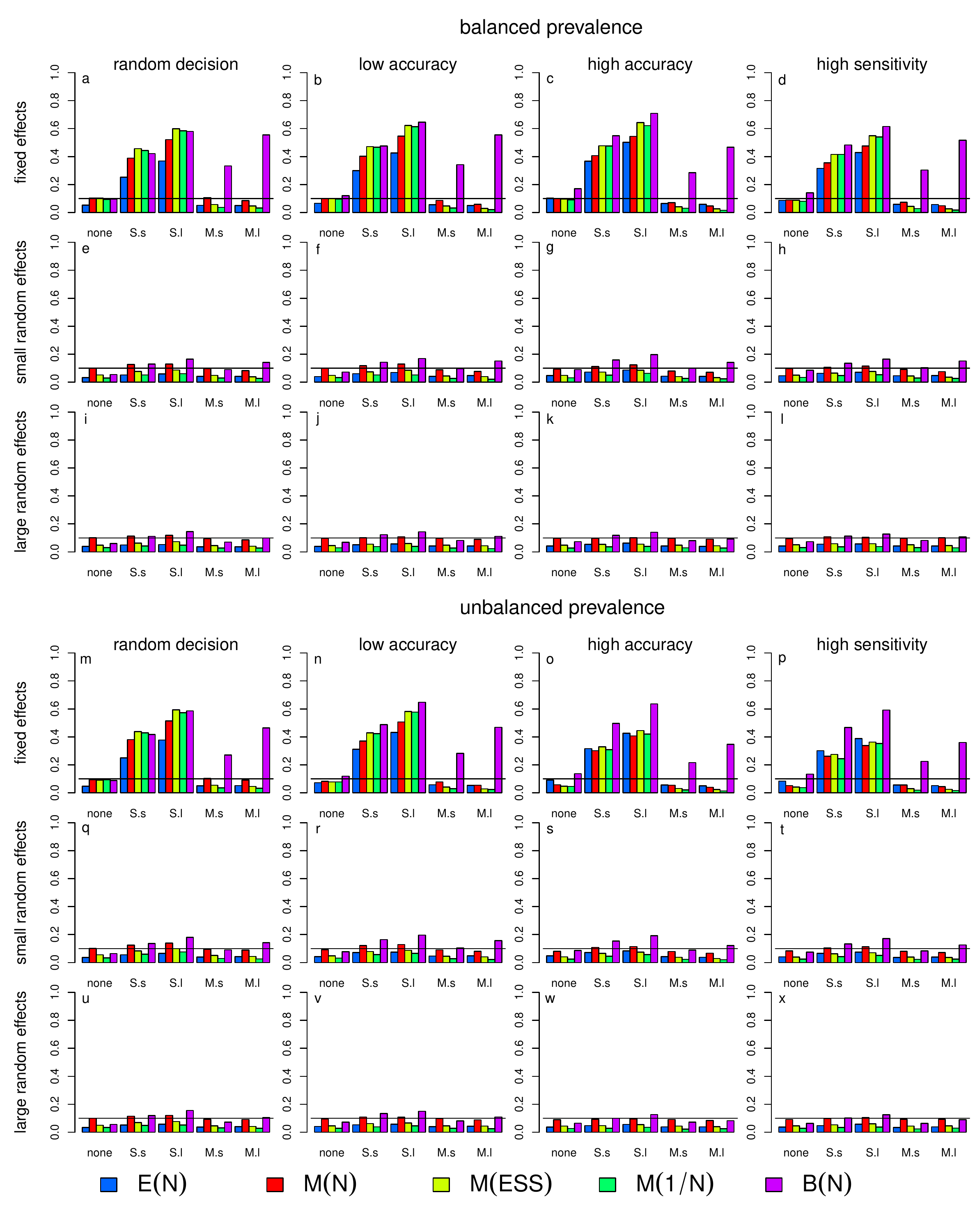}
	\caption[3pt]{Type I error rates and statistical power in case of $k=30$ for linear regression and rank correlation tests using $N$ or $\ESS$ combined with $\lnDOR$. Abbreviations: S.s = selection with small PB, S.l = selection with large PB, M.s = mixture with small PB, M.l = mixture with large PB.}
\label{step2N}
\end{figure}

However, tests based on Egger, Macaskill, or Begg that used the total sample size $N$ or similarly used $\ESS$ or $1/N$ instead of $\SE$ did not suffer from high $\alpha$-inflation. Instead, they appeared more or less conservative even when many studies were included in the meta-analysis (see Figure~\ref{step2N}). In case of fixed effects (see first and fourth row in Figure~\ref{step2N}), all tests had an acceptable amount of power, when PB was simulated by the selection method. Notably, only rank correlation tests were able to identify PB, which was simulated by the mixture approach. Unfortunately, all of these tests generally had low power in case of random effects (rarely above the nominal $\alpha$-level; see second, third, fifth, and sixth row in Figure~\ref{step2N}). The same was true for the arcsine tests of R\"ucker \textit{et al.} \cite{rucker2008}. The prevalence $\pi$ did only have small effects (compare first to third with fourth to sixth row in Figure~\ref{step2N}), at least for the non-extreme values of $\pi$ that were chosen in this study. Taken together, when the investigated diagnostic test is better than a random decision 
 or when random effects such as different cut-off values are present, the methods based on linear regression or rank correlation cannot be recommended for diagnostic meta-analysis, because of inflated $\alpha$-levels or very low power.

\begin{figure}[btp] 
  \centering
    \includegraphics[scale=0.75]{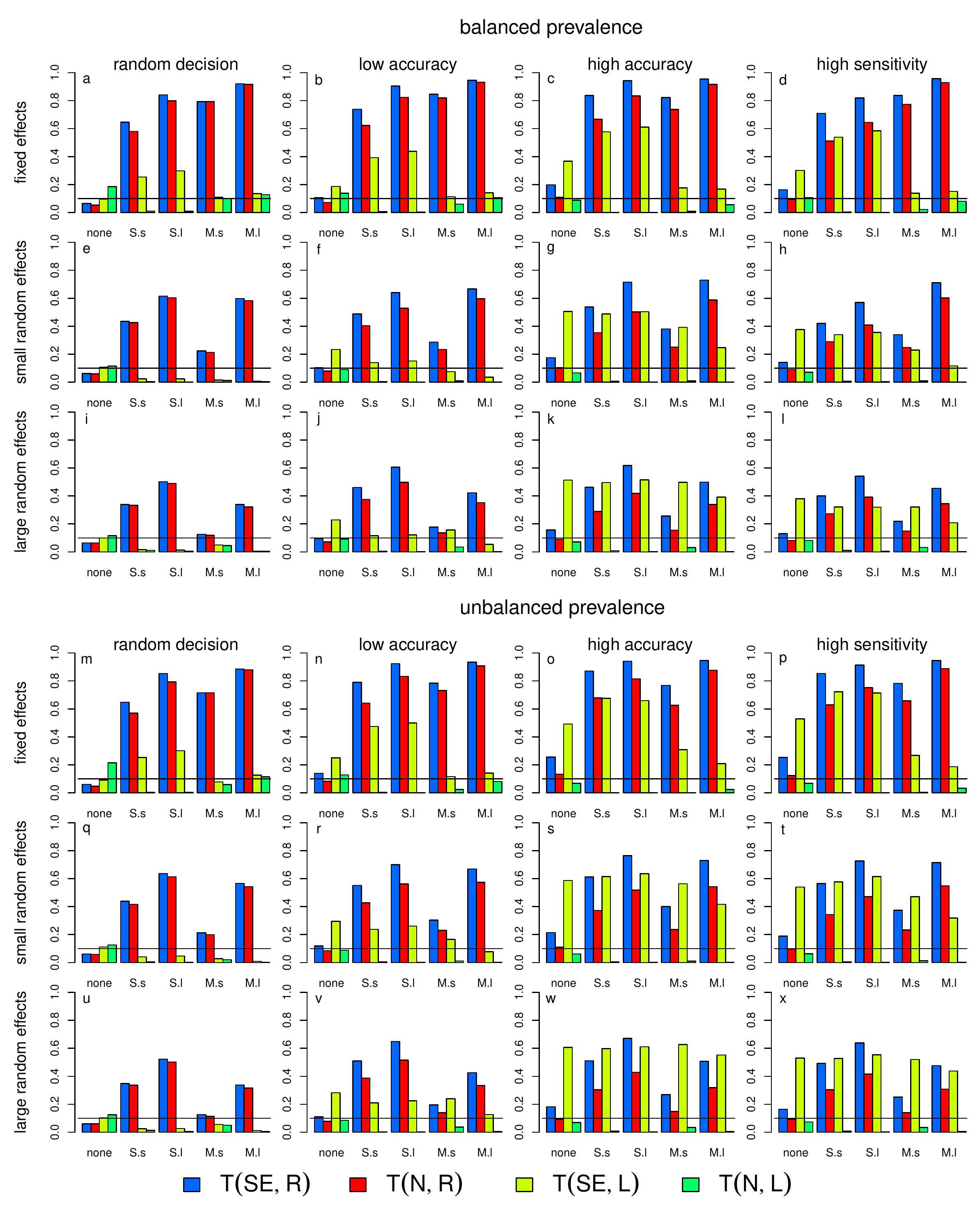}
	\caption[3pt]{Type I error rates and statistical power in case of $k=30$ for trim and fill combined with $\lnDOR$. Abbreviations: S.s = selection with small PB, S.l = selection with large PB, M.s = mixture with small PB, M.l = mixture with large PB.}
\label{step2T}
\end{figure}

\begin{figure}[btp] 
  \centering
    \includegraphics[scale=0.75]{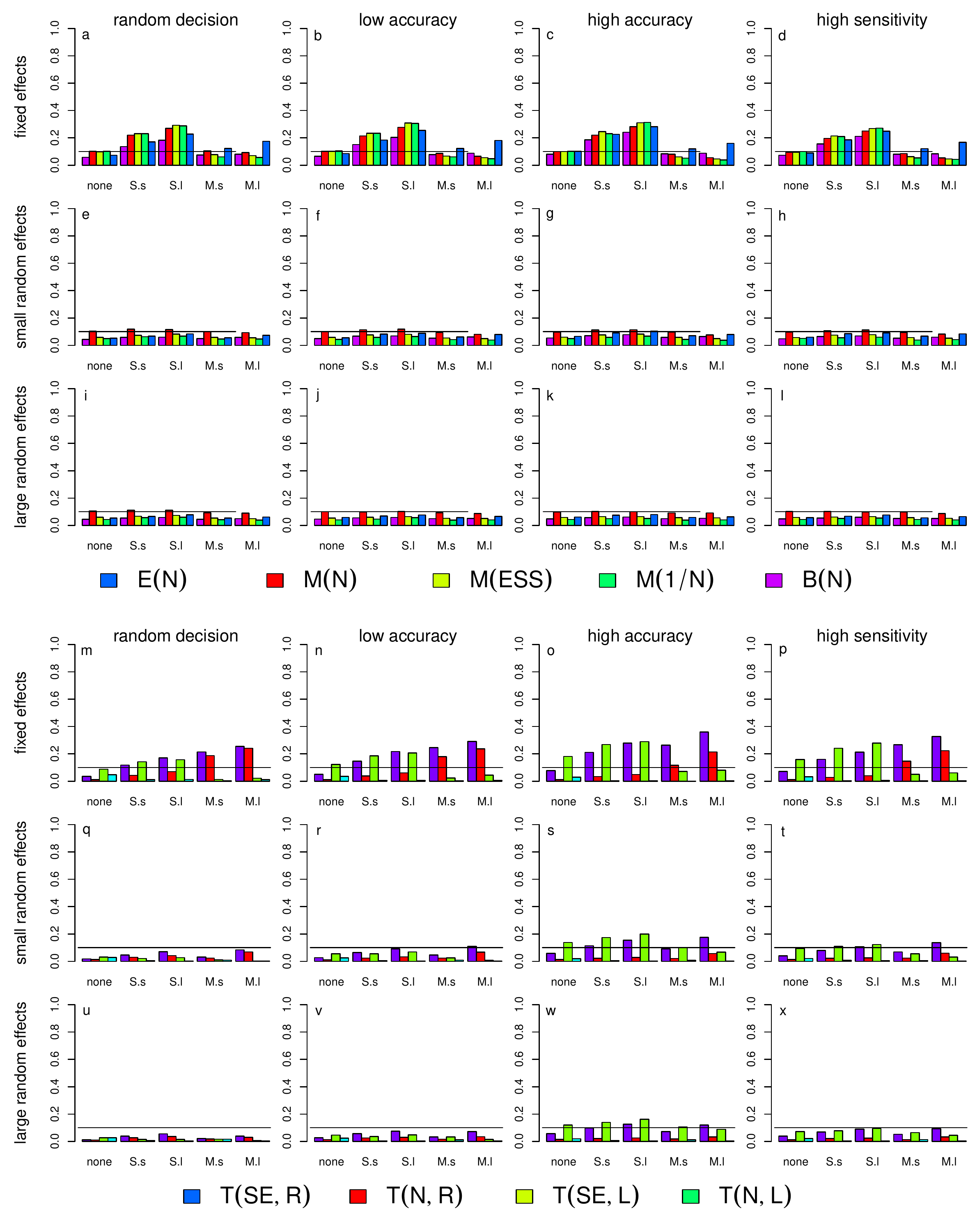}
	\caption[3pt]{Type I error rates and statistical power in case of $k=10$ and balanced prevalence for linear regression and rank correlation tests using $N$ or $\ESS$ as well as trim and fill each combined with $\lnDOR$. Results for $k=10$ and unbalanced prevalence are similar. Abbreviations: S.s = selection with small PB, S.l = selection with large PB, M.s = mixture with small PB, M.l = mixture with large PB.}
\label{step2NT}
\end{figure}

In contrast, trim and fill had non-inflated or only slightly inflated $\alpha$-levels and medium to high power (when $k = 30$; see Figure~\ref{step2T}), at least when the number of missing studies was estimated by (16). Generally, the results of $\T(\SE,R)$ and $\T(N,R)$ were independent of $\pi$ and they were able to detect both types of PB (i.e. mixture and selection). $\T(\SE,R)$ had more power than $\T(N,R)$, but was slightly liberal when the diagnostic test had overall high accuracy or at least high $\Sen$ (see third and fourth column in Figure~\ref{step2T}), whereas $\T(N,R)$ was non-liberal in all cases. Both tests lost power in the presence of random effects. Importantly however, these losses were much smaller compared to all other non-liberal or slightly liberal tests. When only a few studies were included in the meta-analysis ($k = 10$), trim and fill was rather conservative and had quite low power, which was similar to other tests with non-inflated or only slightly inflated $\alpha$-levels (see Figure~\ref{step2NT}). Interestingly, the estimator of the number of missing studies stated in (17) did not work properly for almost all parameter combinations. Among others, this might be due to the fact that (17) tends to underestimate the number of missing studies \cite{duval2000}. 

\section{Discussion}

The present simulations confirmed that statistical tests based on funnel plots are able to detect PB in diagnostic meta-analysis under diverse conditions. Summarizing the results, trim and fill combined with the log diagnostic odds ratio, more precisely $\T(\lnDOR,\SE,R)$ or $\T(\lnDOR,N,R)$, were best to detect PB in diagnostic meta-analysis, although both lacked power when the number of studies per meta-analysis was small. Furthermore, $\T(\lnDOR,N,R)$ was able to hold the nominal $\alpha$-level in all cases and did not have much less power than $\T(\lnDOR,\SE,R)$, even in situations with between study heterogeneity. This finding demonstrates that funnel plots based on the total sample size $N$ (but not on the SE) also provide enough information, so that PB can be detected.    

In contrast to trim and fill, the common tests of Egger, Macaskill, and Begg were too liberal in the presence of random effects or they had very low power. Thus, earlier findings were replicated (for instance see \cite{deeks2005,harbord2006,peters2006}) and it can be concluded that those common tests cannot be recommended for diagnostic meta-analysis. The advantage of the tests proposed by Deeks \textit{et al.} \cite{deeks2005}, which use ESS, was that those tests were mostly able to hold the nominal $\alpha$-level, but at the expense of low power when between study heterogeneity was large. As most diagnostic tests are way better than random decisions and as cut-off values often vary between studies, tests of Egger, Macaskill, and Begg will often be misleading. Accordingly, these tests and their numerous existing variations cannot be recommended in diagnostic meta-analysis. 

Both the $\alpha$-inflation of certain tests that were combined with $\lnDOR$ and $\lnt$ and the odd performance of $Y$ and $K$ when combined with one-sided tests need some explanation. First, the distributions of the UMs, especially the ones of $Y$ and $K$ are more symmetric in the presence of PB than in its absence (see Figure~\ref{densities}). Second, the SEs of all the four applied UMs are dependent on the underlying effect. With $N$ held constant, $\SE(\lnDOR)$ and $\SE(\lnt)$ reach higher values, when the underlying effect differs more from zero. For $\SE(Y)$ and $\SE(K)$ it is contrariwise. For example consider Egger's regression applied to a meta-analysis of a well performing diagnostic test in which PB does not exist. If $\lnDOR$ or $\lnt$ are used, $\UM \cdot \SE^{-1}$ will be overly constant and greater than zero across different values of $\SE^{-1}$, because the precision gets higher when the UM is close to zero. Conclusively, the probability of $b_{0} > 0$ is increased, which results in a liberal test. In contrast, if $Y$ or $K$ are applied, small values of the UMs will be associated with lower precision, so that the probability of $b_{0} < 0$ is increased, which results in a conservative test with very low power (when the hypothesis is that $b_{0} > 0$). Trim and fill combined with $\lnDOR$ seems to compensate for these problems. Therefore, we recommend its application in diagnostic meta-analysis.

Besides the fact that a simulation by design can never display reality exactly (and is therefore always slightly wrong), there are some limitations of our modeling assumptions that have to be discussed in the following. First, we decided to directly sample the true logits of $\Sen$ and $1-\Spe$ from the random effects model of Reitsma \textit{et al.} \cite{reitsma2005}. This approach was different from and probably less intuitive than earlier simulation models \cite{deeks2005}. However, the bivariate model is a common and valid approach for performing diagnostic meta-analysis \cite{macaskill2010,ma2013,leeflang2008} and therefore its usage as a sampling model appears reasonable. Second, it might be more realistic in some diagnostic settings to assume a long-tailed distribution for the sample size $N$ of each study instead of a uniform distribution, although this should only have a minor impact on the results. Third, two distinct methods to introduce PB were applied in the present study with one of them being very different from the methods proposed in literature. Interestingly, the mixture approach to introduce PB yielded similar results to the selection approach although both arose from different theoretical assumptions. Fourth, the PB mechanisms we studied are not exhaustive: A reviewer of the paper pointed out that a selection mechanism based solely on sensitivity would have been a possibility. Fifth, we assumed a perfect gold standard, which seems common in simulations of diagnostic accuracy, but might not be adequate for every diagnostical setting. As the modeling of an imperfect gold standard would have further complicated our simulations and might not have led to very different results, this limitation was considered as acceptable.
  
Sixth and last, as described in Section 1, there are several reasons other than PB (such as different cut-off values, study quality, or heterogeneity of the examined populations) that may cause funnel plot asymmetry. In the present study, these biases were modeled by the covariance matrix of the true logits of $\Sen$ and $1-\Spe$. As indicated by our results, only trim and fill could discriminate adequately between this heterogeneity and PB. In contrast to PB, no other bias did have an explicit \textit{directional} effect on the funnel plot asymmetry in our simulation model. Instead, the effect of the between study heterogeneity largely depended on the respective applied UM and its SE. However, in real diagnostic meta-analysis, other biases may have directional effects on funnel plot asymmetry as well. Unfortunately, the direction is not always known. In meta-analysis of treatment effects, smaller studies often have a poorer methodological quality, which may lead to an overestimation of the effects in these studies. However, this may not be the case in diagnostic meta-analysis: As larger retrospective studies may obtain their test results from large clinical databases and do thus have more heterogeneous and possibly not fully appropriate samples for the respective research question, they may be of poorer quality in contrast to studies with small but more carefully chosen samples \cite{deeks2005}. Depending on possible directions of other biases, the performance of tests to detect PB may vary heavily, which further complicates the detection of PB. 

Importantly however, our results for those tests already investigated in literature turned out to be quite similar to earlier findings \cite{deeks2005,peters2006,harbord2006}, which supports our assumptions and also validates our findings.

In summary, the present study provides evidence that trim and fill combined with log diagnostic odds ratios is superior to other combinations of tests and UMs when testing for PB in diagnostic meta-analysis. Moreover, the tests based on linear regression or rank correlation cannot be recommended for diagnostic meta-analysis, either because of highly inflated $\alpha$-levels or very low power.  

\section*{Acknowledgements}
The work of the second author was funded by DFG Grant HO 1286/7-2.

\begin{center} 
\bibliography{paper}{}
\bibliographystyle{wileyj}
\end{center}



\end{document}